\documentclass[iop,apj,tighten]{emulateapj}
\usepackage{amsmath,amstext}
\usepackage[breaklinks,colorlinks,citecolor=blue,linkcolor=magenta]{hyperref} 
\usepackage[all]{hypcap} 
\usepackage{verbatim}

\usepackage{graphicx}	
\usepackage{amsmath}	
\usepackage{amssymb}	

\begin{document}

\title{A Comparison between Radio Loud and Quiet Gamma-Ray Bursts, and Evidence for a Potential Correlation between Intrinsic Duration and Redshift in the Radio Loud Population.}

\author{Nicole M. Lloyd-Ronning\altaffilmark{1,2}, Ben Gompertz\altaffilmark{3}, Asaf Pe'er\altaffilmark{4,5}, Maria Dainotti\altaffilmark{6,7,8}, Andy Fruchter\altaffilmark{9}\\ }
\affil{$^{1}$Center for Theoretical Astrophysics \& CCS-2, Los Alamos National Laboratory, Los Alamos, NM 87544}
\affil{$^{2}$ University of New Mexico, Los Alamos, Los Alamos, NM 87544}
\affil{$^{3}$ Department of Physics, University of Warwick, Coventry UK CV4 7AL}
\affil{$^{4}$ Department of Physics, University College Cork, Cork, Ireland}
\affil{$^{5}$ Department of Physics, Bar-Ilan University, Ramat-Gan 52900, Israel}
\affil{$^{6}$ Department of Physics \& Astronomy, Stanford University, Stanford, CA USA 94305}
\affil{$^{7}$ Obserwatorium Astronomiczne, Uniwersytet Jagiellonski, Krakow, Poland }
\affil{$^{8} $ INAF,  Via Gobetti 101, Bologna, Italy}
\affil{$^{9} $ Space Telescope Science Institute, Baltimore, MD USA 21218}
\altaffiltext{*}{lloyd-ronning@lanl.gov}

\begin{abstract}

We extend our study of energetic radio loud and quiet gamma-ray bursts (GRBs), suggesting these GRBs potentially come from two separate progenitor systems.  We expand the sample from our previous paper \citep{LRF17} and find our results are strengthened - radio quiet GRBs have significantly shorter intrinsic prompt duration, and are also less energetic on average.  However, the tenuous correlation between isotropic energy and intrinsic duration in the radio dark sample remains tenuous and is slightly weakened by adding more bursts.  Interestingly, we find an anti-correlation between the intrinsic duration and redshift in the radio bright sample {\em but not the radio dark sample}, further supporting that these two samples may come from separate progenitors.  We also find that very high energy ($0.1 - 100$ GeV) extended emission {\em is only present in the radio loud sample}. There is no significant difference between the presence of X-ray/optical plateaus or the average jet opening angles between the two samples.  We explore the interpretation of these results in the context of different progenitor models.  The data are consistent with the radio loud GRBs coming from a Helium-merger system and the radio quiet GRBs coming from a collapsar system, but may also reflect other dichotomies in the inner engine such as a neutron star versus black hole core.

\end{abstract}

\keywords{(stars:) gamma-ray bursts: general
}


\section{Introduction}

 The energetics and timescales of gamma-ray bursts, along with other indicators including locations in their host galaxies and associations with counterparts (e.g. supernovae for the long bursts, and  a gravitational wave counterpart from a double neutron star merger for at least one short burst, \cite{Ab17}) have allowed us to define a general picture of GRB progenitor systems.  Short gamma-ray bursts (with prompt gamma-ray emission lasting less than $\sim 2s$) appear to be associated with older, compact-object mergers.  Long gamma-ray bursts (with prompt emission lasting greater than $\sim 2s$) seem to be associated with younger, massive stars (for reviews summarizing these results see \cite{Pir04, ZM04,Mesz06,GRRF09,Berg14,DAvanz15,KZ15, Lev16} and references therein). \\
 
 However, the nature of the progenitor - particularly for long GRBs - is still under debate.  Promising potential models include black hole-accretion disk systems formed from the collapse of a massive star \citep{Woos93,MW99} or merger systems (e.g. a helium star merging with a black hole, \cite{FW98}). Alternatively, the compact object driving the GRB outflow may not be a black hole-accretion disk system, but instead may consist of a rapidly rotating, highly magnetized neutron star (i.e. a magnetar, \cite{Usov92,Thomp94,Zhang01}).
 
 Observations of the gamma-ray burst afterglow - which carry information about the environment of the progenitor - can potentially shed light on this problem.  For long gamma-ray bursts, about $95 \%$ have X-ray afterglows, about $70 \%$ have optical afterglows, but only about $31 \%$ have a detected radio afterglow. In an attempt to understand the low detection rate of radio afterglows, \cite{HGM13} used visibility stacking techniques to examine 178 long gamma-ray bursts with a total of 737 radio observations.  They concluded that some GRBs without radio afterglows are truly radio dark (in other words, the lack of detection is not just due to a sensitivity limit), and suggested that there exist two populations of GRBs - radio loud and radio quiet. And indeed active galactic nuclei (AGN), which in many ways can be considered scaled cousins of GRBs (e.g. \cite{Nem12,Zhang13,wu16,LR18}), have truly distinct radio loud and quiet sub-populations \citep{WC95,Kell16}.   
   
  Motivated by this, \cite{LRF17} (hereafter LRF) examined a population of ``bright'' long GRBs (in the sense that their bolometric isotropic equivalent gamma-ray energy is $> 10^{52} {\rm ergs}$), with and without radio afterglows using the data of \citet{CF12}.  LRF found that the radio quiet sub-population has significantly shorter average intrinsic prompt duration (i.e. during the gamma-ray burst itself, as opposed to the afterglow), as well as lower isotropic equivalent energy.  They found no significant difference in redshift among the two samples, nor between long GRBs with and without radio afterglows in "dim" bursts (with isotropic equivalent energy $< 10^{52} {\rm erg}$).  Performing a number of other tests, they concluded the radio loud and radio quiet ``bright" bursts show evidence of being from separate populations, potentially originating from different progenitors (or at least progenitors in different environments).\\

 Here, we extend our sample to include long GRBs observed after the \cite{CF12} sample was published.  We find that our previous results hold - radio quiet bursts are significantly shorter in prompt duration and less energetic in terms of isotropic gamma-ray energy. In addition, however, we find that intrinsic duration is anti-correlated with redshift in the radio bright sample, {\em but not in the radio dark sample}. We suggest this further supports that the radio loud and quiet GRBs come from separate populations.  We also compare a number of other properties between the two samples, including whether extended GeV emission was detected in these GRBs.
 
  Our paper is organized as follows.  In \S 2, we describe our data sample. In \S 3, we discuss our statistical analysis and results. In \S 3.1, we report the new correlation found between intrinsic duration and redshift in the radio loud sample only, and we discuss the selection effects that could be affecting this correlation.  In \S 4, we examine the samples' connections to other GRB properties such as the existence of very high energy ($>10 MeV$) emission, the presence of X-ray and optical plateaus and jet opening angle.  An interpretation in terms of progenitor models is given in \S 5 and conclusions are presented in \S 6.

\section{Data}
  Our original data sample was taken from \citet{CF12} - a sample of 304 long GRBs for which radio follow-up observations were performed. The observations were made at frequencies ranging from 0.6 GHz to 666 GHz, with the majority of observations taken at a frequency of 8.5 GHz. The times of the observations ranged from $\sim$ .6hr (.026 days) to 1339 days after the prompt gamma-ray trigger.  Of these follow-up efforts, 95 resulted in detections, 206 in non-detections, and 3 were unconfirmed.  Of the  detections, most peaked in the radio band between 3 and 10 days after the gamma-ray burst trigger.  We refer the reader to \cite{CF12} for the details of the radio observations (see in particular, their Table 3).
  
  We selected those bursts with redshift measurements, so that the intrinsic duration $T_{int}$ could be calculated (i.e. corrected for cosmological time dilation: $T_{int} = T_{90}/(1+z)$, where $T_{90}$ is defined as the time interval over which 90\% of the total observed counts have been detected in the prompt emission), and an estimate of the isotropic emitted energy could be made.  After selecting for the ``bright" bursts with $E_{iso} > 10^{52} {\rm erg}$, we were left with a total of 96 GRBs - 59 with a radio afterglow and 37 with no radio afterglow.

   In this paper, we add to our sample GRBs followed up in the radio since 2012, making the same cuts described above - they must have a redshift measurement and an $E_{iso} > 10^{52} {\rm erg}$. To do so, we utilized Jochen Greiner's GRB page\footnote{ http://www.mpe.mpg.de/~jcg/grbgen.html}, which has information on follow-up of all detected GRBs.  This brings our sample to 78 GRBs with a radio afterglow and 41 without. The distribution of our radio flux densities for the radio bright sample follow that of the total set of GRBs with radio flux measurements (including those with no redshift measurements; this distribution can be seen in Figure 4 of \cite{CF12}). We note that when multi-wavelength follow-up was possible, {\bf all 119} of GRBs in our combined radio loud and quiet samples have measured X-ray and optical afterglows, except for one (GRB150413A was not detected in X-ray). \\

  Several recent studies \citep{Qin13,Lan18}, have pointed out the energy dependence of the bursts' measured duration in gamma-rays, $T_{90}$.  For example, Figure 3 of \citet{Qin13} shows the duration distribution of GRBs for six different gamma-ray detectors. In particular, Fermi's GBM (sensitive at energies of 8-1000keV) shows the peak of the $T_{90}$ distribution shifted to lower values compared to {\em Swift's} BAT (sensitive at energies of 15-150keV). A full analysis of the physical implications behind the spectral energy dependence of $T_{90}$ is beyond the scope of this paper, but we emphasize this point is to be kept in mind in the interpretation of our results. In this paper, we use the $T_{90}$ measurements from the {\em Neil Gehrels Swift} BAT detector.
  


\section{Results}
  We find that when we include long GRBs with radio follow-up since 2012, the results from LRF hold - the intrinsic durations of bright long GRBs without a radio afterglow are significantly shorter than those with a radio afterglow.  In addition, the average isotropic emitted energy in the radio quiet sample is lower than in the radio loud sample.  The average values of intrinsic duration, isotropic energy, and redshift are shown in Table 1, and are consistent with the values found in LRF.

   A Student's $t$-test on the average value of $T_{int}$ (comparing the radio loud and radio quiet samples) gives a .001 probability that they are from the same distribution, statistically strengthening the previous claim of LRF that the intrinsic durations between the radio loud and quiet samples are significantly different (in our previous sample, we found a p-value of .004). We find similar results performing a Student's $t$-test on the average values of $E_{iso}$ between the radio quiet and loud samples.  A Student's $t$-test on the redshift distributions indicates that the radio loud and quiet samples are drawn from the same redshift distribution.

\begin{table}
\centering
	\caption{Average Values of $T_{int}$ and $E_{iso}$ for Bright GRBs}
	\begin{tabular}{lccc} 
    \hline
Sample & $\bar{z}$ & $\bar{T}_{int}$ (s) & $\bar{E}_{iso}(10^{52}$ erg) \\
\hline
No Radio (41 bursts) | & 2.6 & 16. & 9. \\ 
Radio  (78 bursts) | & 2.8 & 39. & 51. \\
\hline
\end{tabular}
\end{table}

\begin{figure*}[!t]
	\includegraphics[width=2.3in]{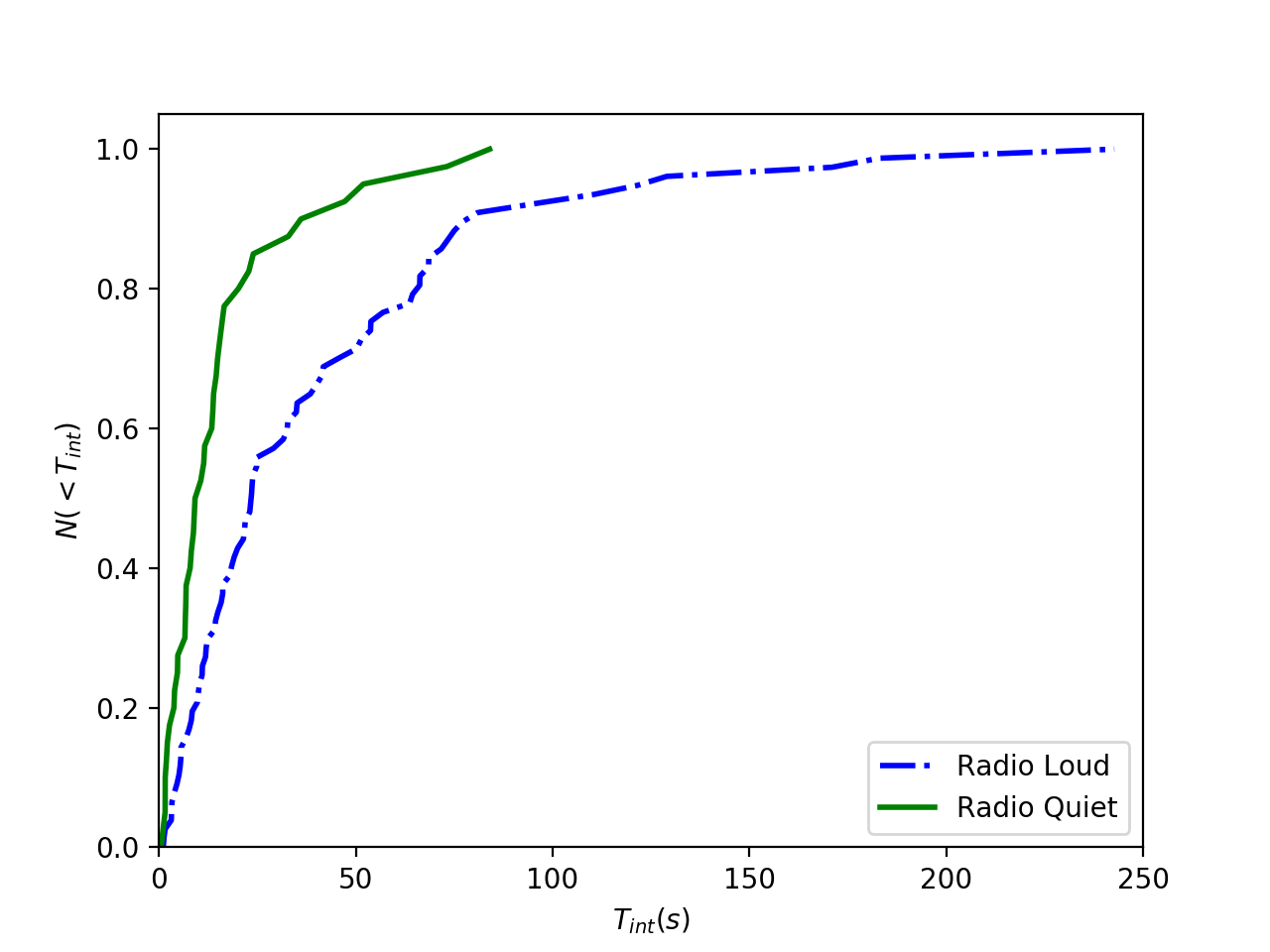}\includegraphics[width=2.3in]{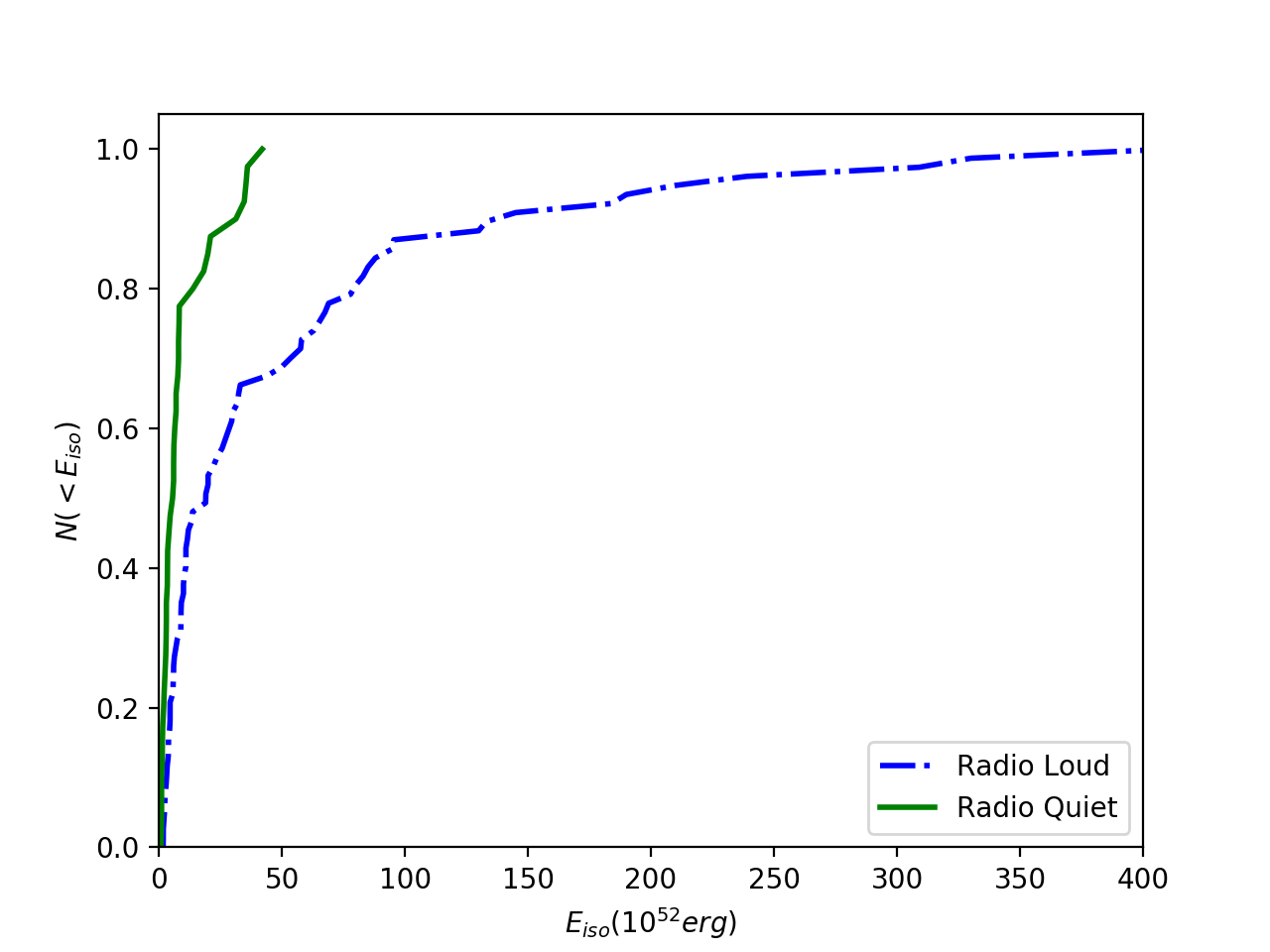}\includegraphics[width=2.3in]{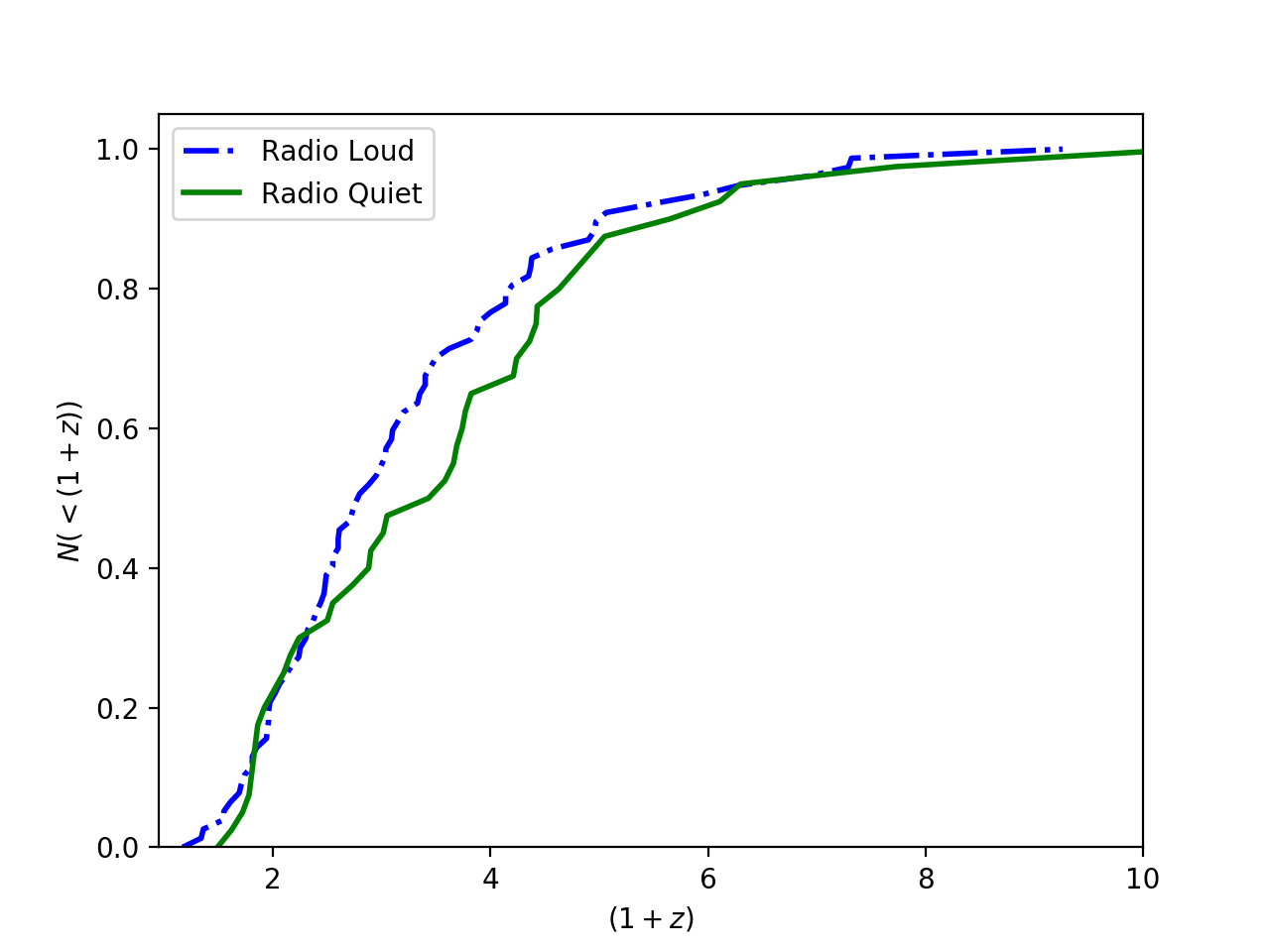}
    \caption{Cumulative distributions of intrinsic duraiton $T_{int}$ (left panel), isotropic equivalent energy $E_{iso}$ (middle panel), and redshift $(1+z)$ (right panel) for the radio bright (dotted blue line) and radio quiet (solid green line) samples.  A KS test indicates the radio loud and quiet samples have significantly different distributions in both $T_{int}$ and $E_{iso}$, but not $(1+z)$.}
    \label{fig:cumdist}
\end{figure*}

\begin{figure*}[!t]
	\includegraphics[width=2.3in]{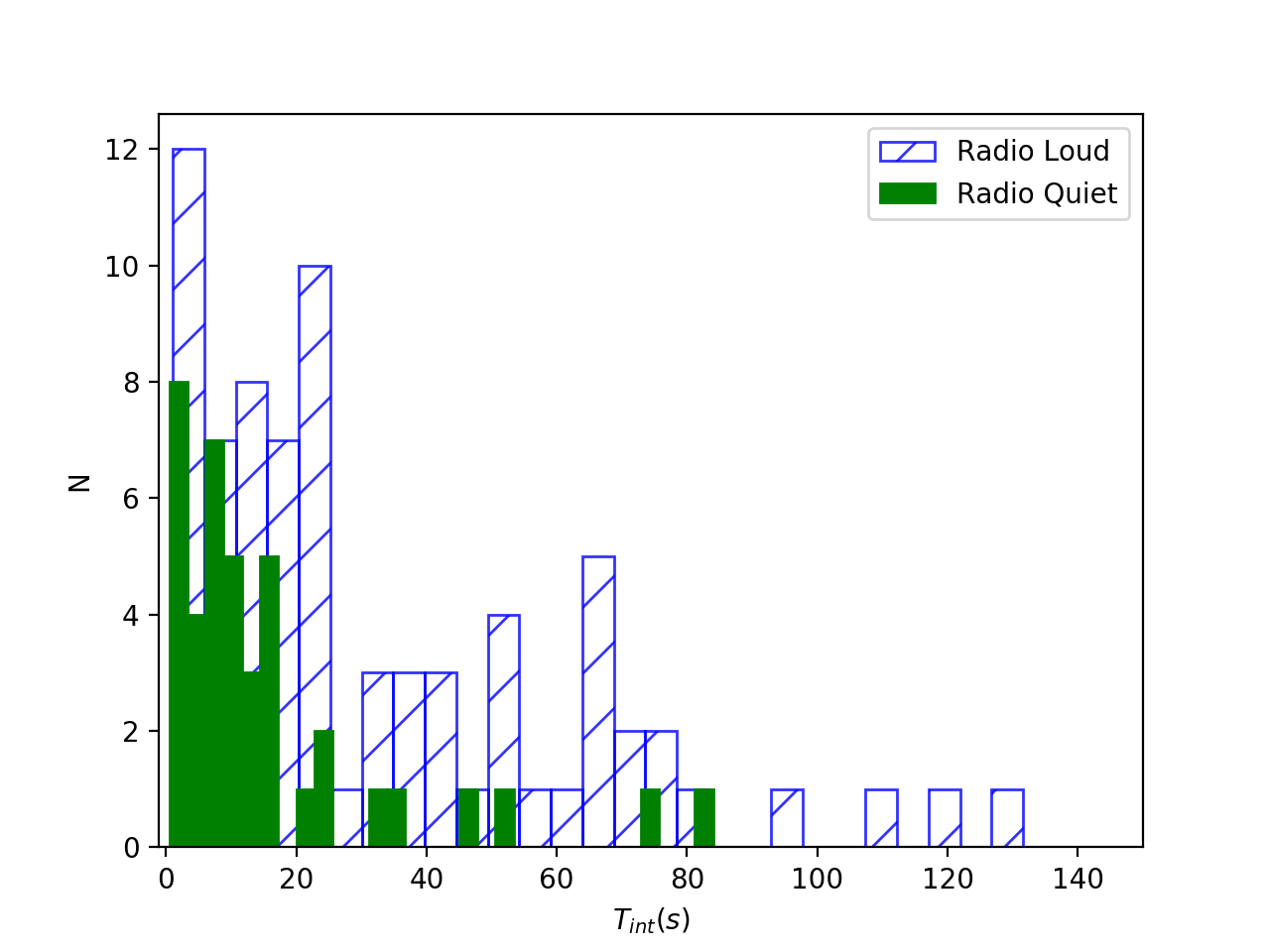}\includegraphics[width=2.3in]{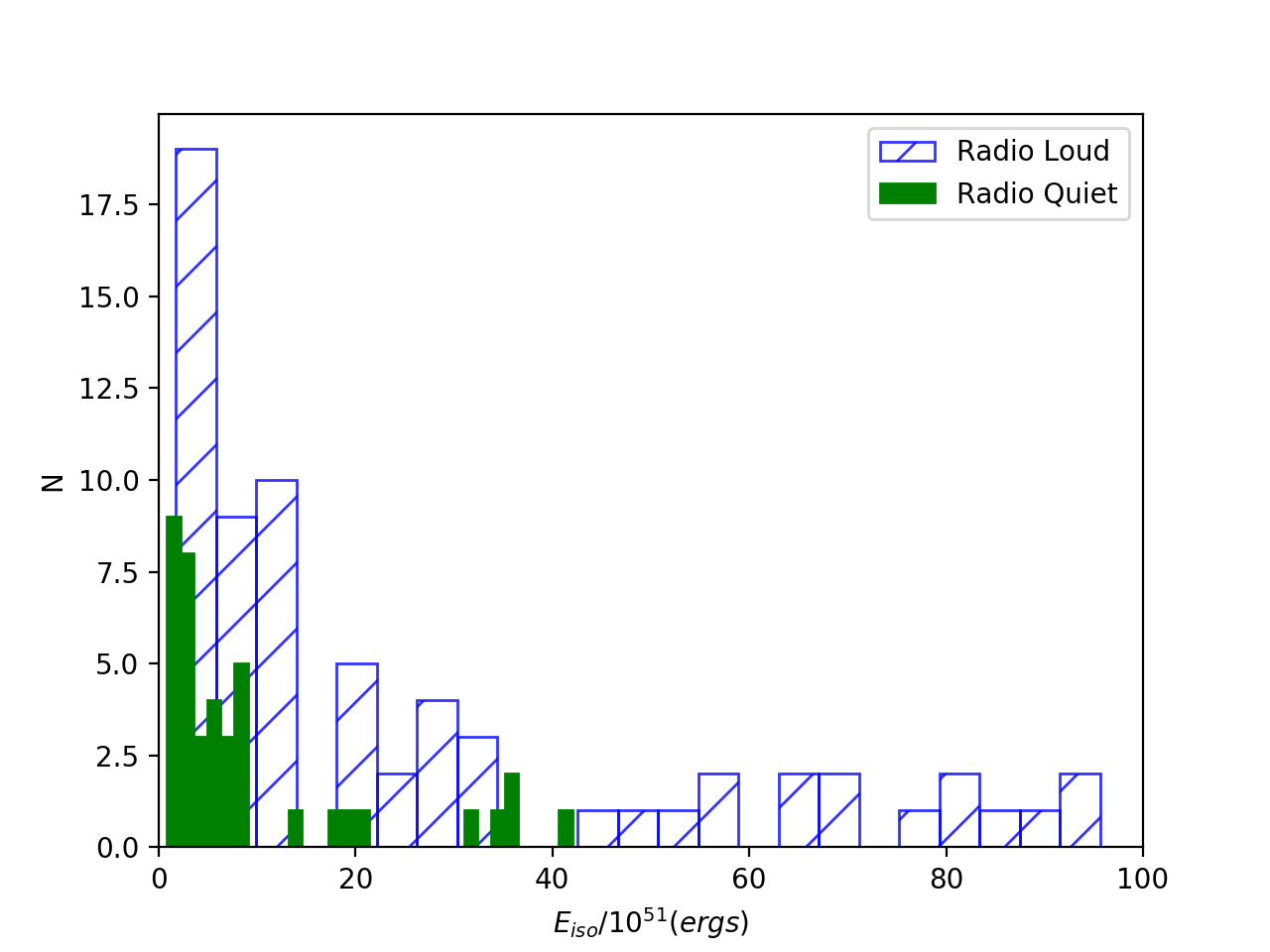}
    \includegraphics[width=2.3in]{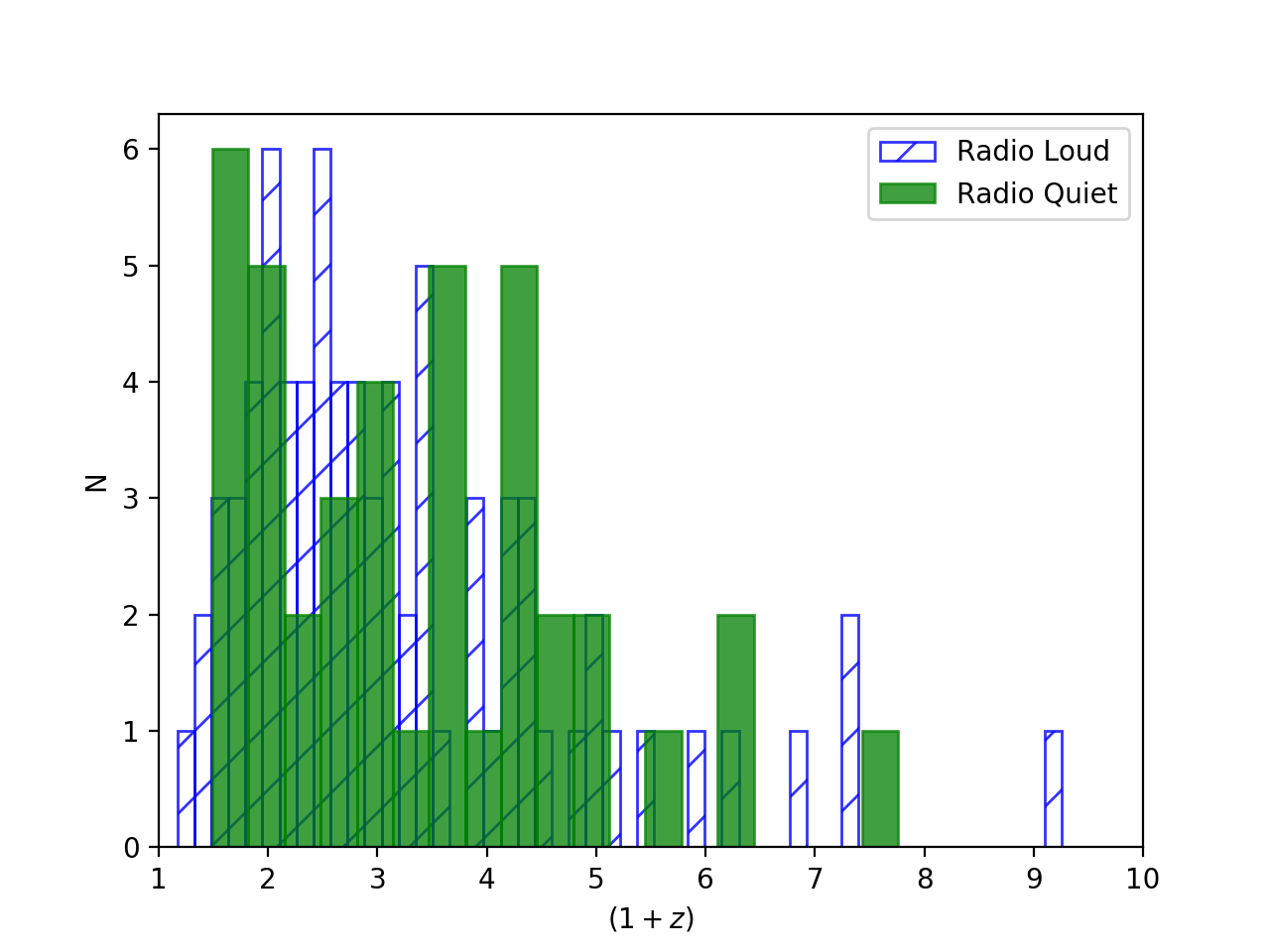}
    \caption{Histograms of the intrinsic prompt gamma-ray durations (left panel), isotropic equivalent energies (middle panel) and redshifts (right panel) of our radio loud (blue) and quiet (green) samples.  A Student's t-test indicates the radio loud and quiet samples have different distributions of $T_{int}$ and $E_{iso}$, but not redshift (1+z).}
    \label{fig:histos}
\end{figure*}
 Figure~\ref{fig:cumdist} shows the cumulative distributions of $T_{int}$, $E_{iso}$, and $(1+z)$ for the radio loud (blue dashed line) and radio quiet (green solid line) samples.  A KS test on the cumulative distributions of $T_{int}$, comparing the radio quiet and loud samples, gives a $2 x 10^{-4}$ probability that they are drawn from the same distribution.  Similarly, a KS test on the distributions of $E_{iso}$ gives a $6x10^{-6}$ probability that $E_{iso}$ comes from the same distribution when comparing the radio loud and quiet samples.  Note that a KS test on their redshift distributions gives a probability value of 0.24 (in other words, no evidence that the redshift distributions are different between the two samples).  Figure~\ref{fig:histos} shows the histogrammed values of $T_{int}$, $E_{iso}$, and $(1+z)$ for the radio loud (blue) and quiet (green) samples. By eye,  differences are evident, but we emphasize the KS test on the cumulative distributions indicates this in a statistically robust way.

  We point out that the GRBs in our sample (with $E_{iso} > 10^{52} erg$) show no correlation between radio flux and $E_{iso}$ (see Figure 5 of LRF) and therefore it is unlikely that the radio dark bursts are simply falling below the radio sensitivity limit.  One may also ask whether the shorter $T_{90}$ measure in the radio quiet sample is due to the measured flux falling below the detector noise, making the burst appear shorter than is. We have checked the difference in the average gamma-ray fluxes between the radio loud and quiet samples and found that although the average gamma-ray flux for the radio-loud sample is slightly higher than the radio quiet sample ($6.2 \ {\rm erg/cm^{2}/s}$ vs. $3.1 \ {\rm erg/cm^{2}/s}$), a Student's t-test shows no significant difference between their distributions.
Therefore, it does not seem that a tip-of-the-iceberg effect is the sole reason behind the difference in the duration of the prompt emission between the radio loud and quiet samples (we discuss this in further detail in \S 3.1 below).
 \\

  We find the tenuous $E_{iso}-T_{int}$ correlation that was present in our previous radio quiet sample (see Figure 4 of LRF) is slightly weakened to $\sim 2.5 \sigma$. This correlation is not (at this point) statistically significant and therefore not a strong discriminator between the potentially different progenitors of these two samples. However, we do note that there is likely to be some amount of crossover between the two samples and it may be that as more bursts are added, this existence (or non-existence) of this correlation will become more apparent.

\subsection{An Anti-correlation between Intrinsic Prompt Duration and Redshift.}
\begin{figure*}[!t]
	\includegraphics[width=3.5in]{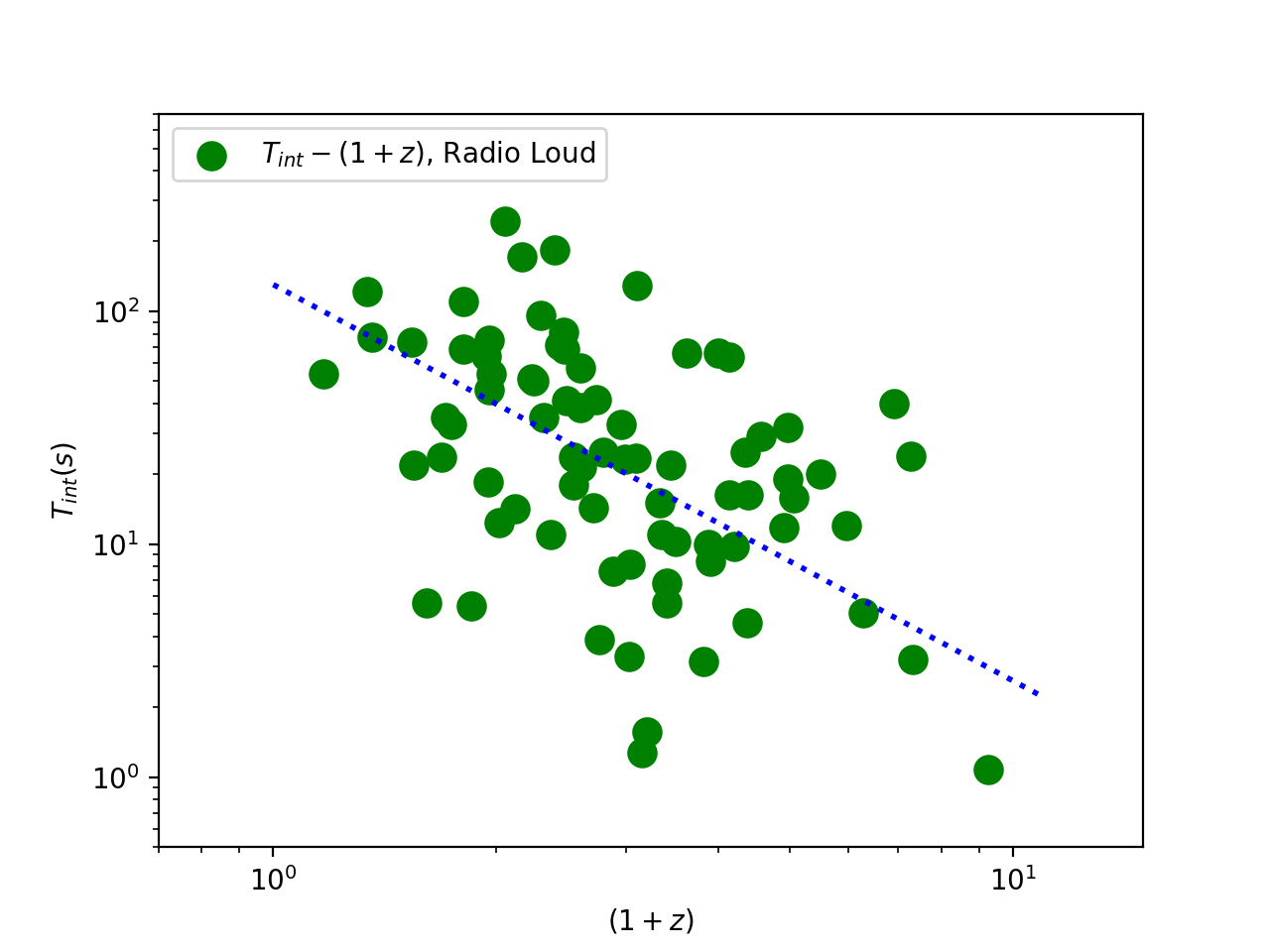}\includegraphics[width=3.5in]{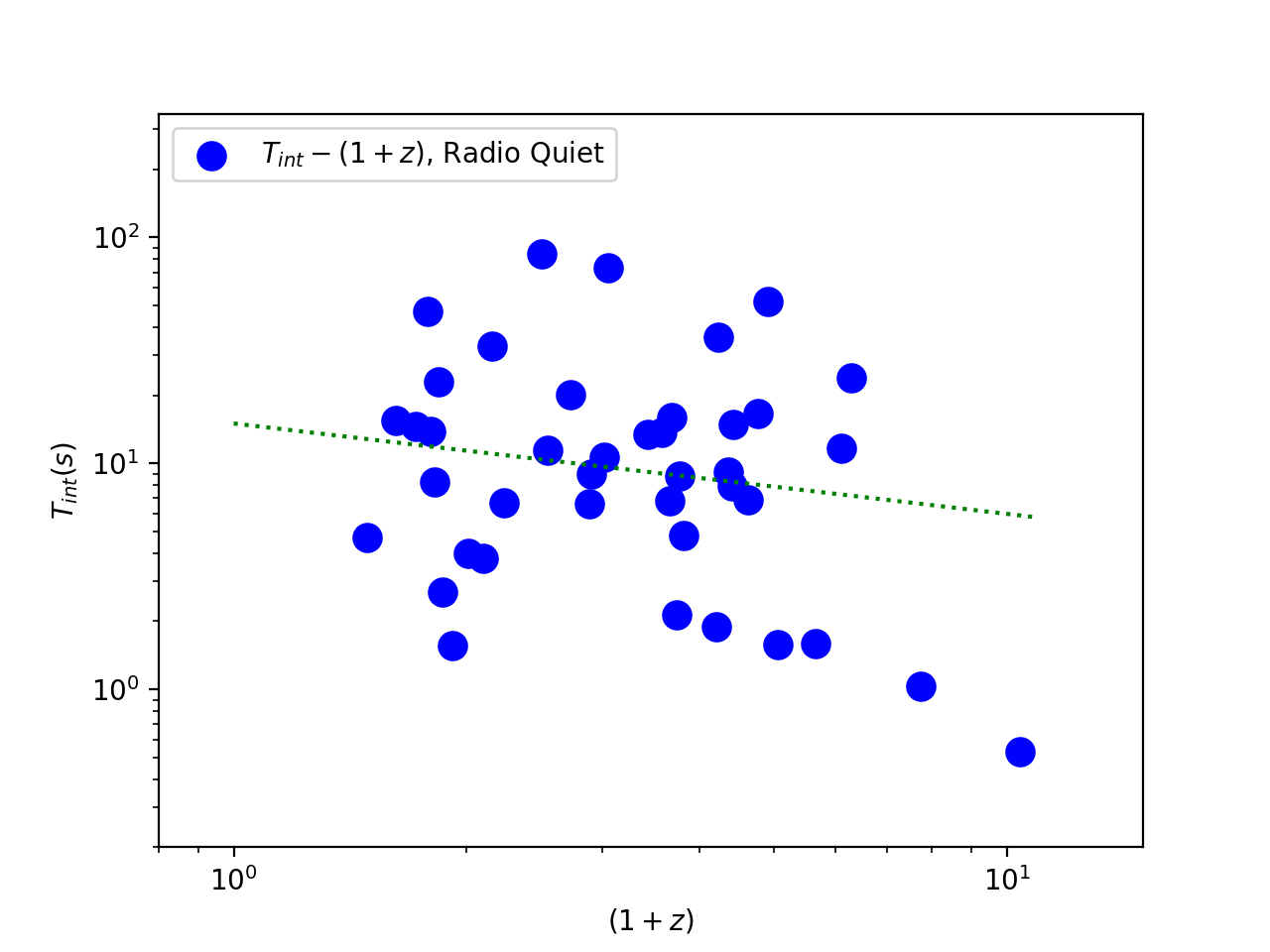}
    \caption{Intrinsic prompt duration vs. redshift for the radio loud (left panel) and radio quiet (right panel) samples.  A statistically significant anti-correlation is present {\em only in the radio loud sample}.  The dotted blue line gives the best parameterization of the correlation in the radio loud sample: $T_{int} \propto (1+z)^{1.4 \pm 0.3}$. In the right panel we show the parameterization of the data, $T_{int} \propto (1+z)^{0.4 \pm 0.5}$, but emphasize {\em there is no statistically significant correlation between these variables for the radio quiet sample.} }
    \label{fig:tintz}
\end{figure*}

  Interestingly, we find a $> 4 \sigma$ anti-correlation between the intrinsic prompt duration and redshift in the radio loud sample, but {\em not} in the radio quiet sample (the p-value from a Kendell's $\tau$ test on $T_{int}$ and $(1+z)$ is $p=3x10^{-5}$ for the radio loud sample, and $p=0.3$ for the radio quiet sample). The data are shown in Figure ~\ref{fig:tintz}.  The functional form of the correlation in the radio bright sample can be parameterized roughly as $T_{int} \propto (1+z)^{1.4 \pm 0.3}$, and is shown by the dotted blue line in the left panel of Figure ~\ref{fig:tintz}.  Again, there is no statistically significant correlation in the radio quiet sample; the best fit parameterization of the radio quiet data is $T_{int} \propto (1+z)^{0.4 \pm 0.5}$.  \\

   It is important to consider whether there are any biases in the $T_{int}-(1+z)$ plane that could potentially lead to an artificial correlation between these two variables. For this to be the case, there would need to be a bias against detecting short, nearby GRBs and/or long, distant GRBs.  In addition, if such a selection effect is artificially producing the correlation, this bias would need to be present {\em only} in the radio loud sample, but not in the radio quiet sample (since the radio quiet sample shows no anti-correlation between $T_{int}$ and $(1+z)$). We stress again that the redshift distributions of the two samples are statistically similar.
   
    As stated earlier, there is a potential concern that bursts at higher redshifts will have some of their flux shifted below the detector noise and may therefore appear shorter-lived than they really are.  As a result, a burst of a given luminosity at higher redshift will have a measured $T_{90}$ that is less than a burst of the same luminosity at a lower redshift (note that cosmological time dilation will offset this affect to some degree).  Additionally, the shape of the pulse can play a significant role in this effect. For a top-hat shaped pulse, the intrinsic duration will not be affected as long as the flux is above the noise.  For a triangle-shaped pulse, however, the measured duration may appear shorter at higher redshifts because of more of the pulse falling below the noise.  The competing effect is, of course, cosmological time dilation which will lengthen the observed duration, and so it is important to understand how the shape of the pulse plays a role in the measured duration. 
   
   We note \cite{LJ13} looked at this effect in detail.  They artificially shifted a sample of GRBs to high redshifts and, convolving with the {\em  Swift} BAT detector response function, examined the new (redshifted) measured $T_{90}$.  They found in most cases, cosmological time dilation played the largest role, so that higher redshift bursts were on average longer.  However, in some cases (depending on the GRB light curve), the effect of the redshifted flux dropping below the detector limit caused bursts at higher redshifts to have a smaller $T_{90}$ (see also \cite{KP13} who examined this effect in simulated GRB light curves).  A few of these bursts that exhibited this latter trend overlap with our radio bright sample.  However, the severity of the effect differs depending on the method they use for their light curve modeling and detector response convolution, and tends to show up mostly (but not always) around redshifts of $(1+z) \sim 4$ or higher.  They conclude from their study that the short durations of high redshift GRBs are not explained simply by band shifting and sensitivity concerns.
   
   There are two reasons this potential bias may {\em not} be playing a role in our analysis: The first is that both of our samples are distributed over the same redshift range so it should be equally present in both samples, whereas the anti-correlation is only present in the radio loud sample. Second, because GRBs have a broad distribution in their luminosities, the severity of this effect can be washed out.  However, a detailed examination of the pulse structure of each burst in our sample is needed to really get a handle on how this effect contributes to the correlation, which we defer to a future publication.\\
   
  Another potential concern is a bias against low intrinsic duration bursts at low redshifts.  This could come about from the detector's time resolution limit. We have examined this effect, evaluating the truncation in the T-(1+z) plane due to the detector time resolution (an estimate of this can be the shortest variability timescale or shortest duration burst detected by {\em Swift}). Using a very conservative value of 0.5 seconds for the time resolution of {\em Swift}, we found the observational truncation produced by this limit has no effect on the correlation (in reality, the time resolution is much smaller than this and makes the observational truncation even less severe).  \\
  
   Our main emphasis here is that - even if this correlation is influenced by observational biases - {\em it is significant that it is only present in the radio loud sample and not in the radio quiet sample}, given that both samples cover a similar redshift range.  In fact, even removing the highest duration bursts ($T_{int}>90s$) in the radio loud sample (so that the radio loud and quiet sample span roughly the same intrinsic duration range), we {\em still} find a statistically significant ($\sim 4 \sigma$) anti-correlation between $T_{int}$ and $(1+z)$ in the radio loud sample only. We discuss the possible physical meaning of this potential correlation in our discussion section (\S 5) below.  However, first we examine each sample's relation to other GRB properties.
  
  

\section{Correlation with Other GRB Properties}
  In this section we analyze whether our radio loud and quiet samples show correlations with the presence of other GRB properties - very high energy emission, plateaus in the X-ray and optical afterglow light curves, and jet opening angles.
 
 \subsection{Presence of Very High Energy Emission}
   Using the data from \cite{Nava18}, we examined bursts in our sample that were detected by the LAT on board Fermi, indicating emission in the $0.1$ to $100$ GeV range.  We found $9$ GRBs in our sample with very high energy emission; of these, $8$ were from our radio loud sample and only $1$ from the radio quiet sample.  Furthermore, the $8$ from the radio bright sample all showed extended emission in this energy range (defined by lasting more than three times the prompt $T_{90}$ emission), while the one in the radio quiet sample did not.
   
   Interestingly, these $8$ radio loud GRBs with very high energy extended emission have an average $E_{iso}/10^{52} {\rm erg}$ of $168$, and an average intrinsic duration of $T_{int} = 62$ s, indicating they are among the brightest, longest-lasting GRBs - perhaps extremes on the progenitor spectrum (see Discussion section below).  Note that Figure 4 of \cite{Nava18} shows that GRBs with LAT emission are in general brighter (in terms of isotropic energy) than the average.
   
   We also point out that, recently, \cite{Ferm18} investigated the nature of the long-lived very high energy emission in GRBs.  Using a combined analysis of the {\em Neil Gehrels Swift} XRT and Fermi LAT emission, they concluded that those with long-lived high energy emission could be a result of synchrotron emission in a {\em wind-like} medium (their arguments are based on the estimated evolution of the synchrotron cooling break in a wind-like medium in the context of a simple synchrotron shock model, e.g. \cite{CL00}). 

 \subsection{Presence of X-ray and Optical Plateaus}

  Many GRBs exhibit plateaus in their X-ray and optical afterglow light curves \citep{Obri06,PV08,Dain10,Zan13,Dain08,Dain13,Dain15,Dain16,Dain17}.  Although it is not always straightforward to determine whether a "plateau" exists, these GRB light curves show some amount of flattening in the temporal decay of their flux over a given energy band, and the degree of this flattening can be analytically defined.

  
 We follow the criterion adopted in \cite{Dain17c} in which the light curves that fulfill the phenomenological \cite{Will07} model are considered as possessing plateau emission in their X-ray light curves. Using their data, we found that 27 out of our 78 bursts in the radio loud sample have X-ray plateaus (with 6 present in their "gold" sample), and that 18 out of 41 bursts in the radio quiet sample have X-ray plateaus (with 5 present in their gold sample).  
 
 This suggests that the presence of a plateau (using their definition) is not a strong discriminator between the radio loud and quiet samples.  This, in turn, could mean there is no difference between their progenitors or it could indicate that - if the radio loud and quiet do indeed come from distinct progenitors - different progenitors can produce plateau phases (e.g. through a magnetar inner engine \citep{Usov92,Thomp94} which may be at least temporarily produced in a wide range of progenitor models; \cite{SD18,Row14,Rea15}).  
 
 Similarly, \cite{Si18} examined the presence of optical plateaus in long GRB light curves.  We compared the bursts in our radio loud and quiet samples to their sample of GRBs with optical plateaus.  We found 10 of the 41 radio quiet bursts have optical plateaus and 12 of the 78 radio loud bursts have optical plateaus.  Again, the presence of an optical plateau does not appear to be a discriminator between our samples, just as in the case of the X-ray plateaus.  

 
 \subsection{Jet Opening Angles}

 When the data exist, we have examined the jet opening angles of both the radio loud and quiet samples.  As with most GRB physical parameters, there is some ambiguity in the way jet opening angles are estimated and there is not a unique, large sample of GRBs with reliable jet opening angles.  We examine the average jet opening angle of the radio loud and quiet samples using three different data sets:
 \begin{itemize}
\item{The first set contain jet opening angles derived using the method described  in the appendix of \citet{Dain17b} and in \cite{Pesc16}.
They assume that the relationship between spectral peak energy and isotropic energy ($E_{p}-E_{iso}$; \cite{LPM00,Am02}), as well as beaming-corrected gamma-ray emitted energy ($E_{p}-E_{\gamma}$) relations \citep{Ghir04} are valid universally at any redshift.  They parameterize these relationships by the form:
$E_{iso}=k_{a}*E_{p}^{\alpha_{a}}$ and 
$E_{iso}(1-cos(\theta_{j}))=k_{g}*E_{p}^{\alpha_{g}}$, where $\alpha_{a}$, $\alpha_{g}$, $k_{a}$ and $k_{g}$ are fitted parameters. Given the GRB redshift, they compute the unknown $cos(\theta_{j})$ through these 2 equations.
  Using these methods, the average jet opening angle for the radio loud sample is $<\theta_{j}> = 7.1 \deg$, and for the radio quiet sample $<\theta_{j}> = 6.5 \deg$.}
\item{Secondly, we used the data from \citet{Ghir04}, who used afterglow light curve jet breaks to estimate the GRB jet opening angle. From this sample, we found jet opening angles for 12 of our radio loud GRBs and only 2 of our radio dark GRBs. We find an average jet opening angle of $<\theta_{j}> = 4.8 \deg$ for the radio loud sample, and $<\theta_{j}> = 5.4 \deg$ for the radio quiet sample (where, again, there are only 2 bursts in this latter ``sample").}
\item{Finally, we used data from \citet{Wang18}, who examined  a sample of GRBs that indicated jet breaks in their optical light curves, and that were consistent with an achromatic break in the X-ray light curve at the same time (the achromatic nature of the break is indicative of a physical jet, as opposed to spectral-energy-dependent evolution).  From their sample, 21 overlapped our radio bright sample and 8 overlapped our radio quiet sample.  We found the average jet opening angle in the radio loud and quiet sample using the \cite{Wang18} data is $<\theta_{j}> = 2.6 \deg$, and $<\theta_{j}> = 2.04 \deg$, respectively.}
\end{itemize}

  In general, the calculation of jet breaks is fraught with uncertainty (see the \citet{Wang18} paper for further discussion of this). For example, when the jet break occurs depends strongly on the assumption of the density profile of the external medium, as well as the velocity profile of the jet (it may have both radial and angular velocity structure). Because of this - combined with the fact that we see no significant difference in the average jet opening angle from either of the three data sets mentioned above -  we conclude that the jet break data is not a reliable distinguishing factor in understanding any potential underlying difference between our radio loud and quiet samples.\\


  


One may also ask how viewing angle (the angle between the observer and the core of the GRB jet) plays a role in our results.  This is yet another parameter in fitting/understanding afterglow light curves, and is degenerate with the other micro and macro physical parameters \citep{MRW98,RLR02,Gran02,Mat18}. The viewing angle, however, will have an affect on the prompt and afterglow light curves (albeit differently in different wavelength bands, depending on the Lorentz factor of the outflow at the time of the observations and keeping in mind the microphysical parameters could vary with time).  We note that our GRBs were selected for being intrinsically energetic in gamma-rays.  And - again - 118 of the 119 GRBs in {\em both} radio loud and quiet samples have measured early-time X-ray and optical afterglows. It is  therefore likely that the bursts in our sample are all ``on-axis" events, and viewing angle/off-axis effects are not playing a dominant role in our samples.\\


\section{Discussion}
  In order to understand our results - that radio dark GRBs tend to be significantly shorter in prompt duration than radio loud GRBs -  we need to understand the nature of the prompt gamma-ray burst duration, as well as the emission mechanism behind the radio afterglow. There are many factors that affect both observed quantities.   Although the radio afterglow is a function of $E_{iso}$, we do not expect (in most standard models of GRB prompt and afterglow emission) that the radio afterglow should depend on the prompt duration.  Hence, our results suggest that the radio loud and dark GRBs may potentially come from two separate progenitors.  To understand our results, we need to address two fundamental questions:
  
  {\em 1) What is the meaning of gamma-ray burst intrinsic duration?}  This parameter can reflect both the amount of matter available to form a disk around the inner engine, as well as the amount of angular momentum in a system (with higher angular momentum leading to longer durations; see, e.g., \cite{JP08} and the recent paper by \cite{Ag18}, who explored angular momentum in a variety of GRB progenitor scenarios).  External environment can also affect the observed gamma-ray duration, even if the gamma-rays are assumed to come from internal shocks (as opposed to the external shock with the surrounding medium). For example, \citet{GM15} showed that $T_{90}$ can be larger than the central engine timescale by a factor of 2 to 3 due to refreshed shocks or late internal collisions.
  
  {\em 2) What is the nature of the radio afterglow emission?} The afterglow is well-modeled by synchrotron emission from an external blast wave, but can come from either the forward shock or reverse shock, and can be affected by self-absorption (which will suppress the radio emission).  Long-lived radio emission from a reverse shock is only expected in cases of low external medium density, so that the radio emission is in a so-called ``slow-cooling" regime (see, e.g., \cite{Las13, Las16, Las18}). The brightness of the radio emission in both the forward and reverse shock will depend on external density, but is also largely affected by microphysical parameters (like fraction of energy in the electrons and magnetic field).
 

\subsection{Distinguishing Progenitors}
If we focus on the intrinsic duration, we may be led to associate our radio quiet sample with the collapsar model \cite{Woos93,MW99}, which produces shorter duration GRBs due to lower angular momentum in the system (relative to a Helium-merger binary system, for example; see, however, \cite{JP08} who showed that collapsar progenitors can produce long GRBs lasting up to about 400s). The lower isotropic equivalent energies could reflect a weaker jet, which could be produced either through neutrino annihilation \citep{Eich89} or the Blandford-Znajek mechanism \citep{bz77}. The absence of the radio afterglow could reflect a dense wind medium self-absorbing the radio spectrum, or - if the radio afterglow is primarily a result of the reverse shock emission - the lack of a long-lived reverse shock in this circumstellar environment.

In this scenario, the radio loud GRBs would come from a He-merger (or similar type) system, which have large angular momentum and produce longer duration GRBs \citep{FW98,ZF00}.  He-mergers can also produce a wide range of energies, and may indeed produce more energetic GRBs than collapsars \citep{Fry13}.  The presence of the radio afterglow may reflect either an optically thin (to radio) circumstellar environment, or the ability of this environment to allow for longer-lived (i.e. slow cooling) reverse shock emission in the radio.  Because He-merger models can in principle have a range of circumstellar environments, this model can accommodate both ISM-like and wind-like circumstellar density profiles \citep{FW98,ZF00}. \\



We note that \cite{GFP18} performed fits to a sample of GRB afterglows, examining whether a wind or ISM-like medium fits best.  We examined the number of GRBs in our sample that overlapped with theirs, and found 13 overlaps in our radio bright sample and 2 overlap in our radio dark sample.  For the 13 in our radio bright sample to which \cite{GFP18} had fits, 7 were wind-like circumstellar profiles, 5 were ISM-like circumstellar profiles and 1 was unknown. For the two bursts from our radio dark sample that overlapped with the \cite{GFP18} fits, one was fit by a wind-like medium and one by an ISM-like medium. We also found that there was no difference in the best fit probabilities, electron energy index, or whether the cooling break was above or below the X-ray band, in the radio loud and quiet samples that overlapped with the \cite{GFP18} fits.

  The radio loud bursts that overlap with the \cite{GFP18} sample are consistent with the picture we described above in which they arise from a Helium-merger system (which can in principle have both ISM-like and wind-like circumstellar profiles).  However, the numbers are admittedly very small and we caution against placing too much weight on the fitted density profiles due to the degeneracy of the parameters that are used in the fits. Therefore, we are left to conclude that - at this point - there appear to be no clear trends between the values of the GRB parameters \cite{GFP18} find in their fits and our radio loud and quiet samples.  

 We also note that the radio flux will have a different dependence on $E_{iso}$ depending on whether it is in a wind or ISM-like external medium \citep{SPN98,CL00}. The exact dependence of the radio flux on $E_{iso}$ depends on the ordering of the spectral break frequencies (for a discussion of these frequencies in the context of synchrotron emission, see \cite{LP02,GS02}); in general, the flux has a weaker dependence on $E_{iso}$ in a wind-like medium (see, e.g., Tables 1 and 2 of \cite{GS02}). 
  
 As we note in \S 3 and show in Figure 5 of LRF, the radio bright data show {\em no} correlation between radio peak flux and $E_{iso}$.  Although this result (a weak correlation between radio flux and $E_{iso}$) is expected in an wind-like medium, we emphasize that the scatter in the other GRB parameters -  the fraction of energy in the magnetic field $\epsilon_{B}$, fraction of energy in the electrons $\epsilon_{e}$, external density, etc. - can wash this correlation/relationship out in either scenario (wind-like or ISM-like).  Therefore, we do not draw any firm conclusions from the presence or absence of a radio flux - $E_{iso}$ correlation until we have a better handle on the values of the other physical parameters that play a role in the GRB spectrum and light curve.  \\

We reiterate - as shown in LRF - that there is no apparent difference between the radio loud and quiet sample with respect to positions in their host galaxies according to the data of \cite{BBF16}.  In addition, both radio loud and quiet GRBs have supernova associations (with supernovae being expected in both collapsar and He-merger models).\\


  It is interesting to compare our findings to what has been discovered in supermassive black hole-accretion disk systems (i.e. AGN).  In the case of AGN, the nature of the radio loud and quiet populations is still under debate; one model that appears to explain the data well \citep{WC95,Chia15} is that the radio loud population is a system that has undergone a merger and has the angular momentum to power a strong radio jet via a Blandford-Znajek mechanism.  This is somewhat analogous to the picture we have suggested above for radio loud GRBs (resulting from a He-merger-type system).\\
 
  Finally, we point out that up to now we have discussed our results in the context of distinguishing between the progenitor system before collapse (He-merger vs. collapsar).  It is possible that the difference between two potential populations of GRBs is a result of a different compact object at the core of the outflow - i.e. a neutron star versus a black hole (regardless of the initial progenitor).  A detailed comparison of the predicted properties of a black hole vs. neutron star central engine is beyond the scope of this paper; however, for further discussion of the properties of neutron star/magnetar central engines, see \cite{Mazz14,Metz15}. 
 
 \subsection{Meaning of the Intrinsic Duration-Redshift Correlation}
 

We would like to understand the meaning of the apparent $T_{int}-(1+z)$ anti-correlation present in the radio loud sample (under the assumption that this is indeed a true correlation and not just a selection effect). If real, this correlation would reflect a true underlying cosmological evolution of the progenitors of radio loud GRBs.  Hence, if these bursts really do indeed come from a He-merger progenitor, we must understand why intrinsic duration would anti-correlate with redshift in these systems. We note there is {\em not} a corresponding $E_{iso}-(1+z)$ correlation in this sample.
  
    At this point, we can only speculate that perhaps higher redshift systems have less angular momentum and/or have less material available for the GRB disk. Presumably Population III stars - if they do indeed create GRBs - also contribute to this correlation, notably at the high redshift end (see \cite{TYB16} for a recent review on the observational signatures of Pop III stars as GRB progenitors).  
Ideally we would like to investigate this question in detail, with robust cosmological simulations of both Helium merger and collapsar systems, including the contribution from Pop III stars.

\section{Conclusions}
The main results of our paper are:
\begin{itemize}
\item{When adding observations since 2012, we confirm the results of LRF: energetic GRBs ($E_{iso} > 10^{52} erg$) with radio afterglows are significantly longer in prompt duration relative to energetic GRBs with no radio afterglows. The average intrinsic prompt gamma-ray duration is $T_{int} \sim 40s$ for radio bright GRBs and $T_{int} \sim 16s$ for radio quiet GRBs.}

\item{In this sample of radio loud and quiet energetic GRBs, radio loud GRBs have significantly higher average isotropic emitted energy $E_{iso}$ relative to radio quiet GRBs ($5.1\times 10^{53} erg$ vs. $9 \times 10^{52} erg$).}

\item{In the {\em radio loud sample only}, there is a significant ($> 4 \sigma$) anti-correlation between intrinsic prompt duration and redshift, with the correlation parameterized as $T \sim (1+z)^{1.4 \pm 0.3}$.  The correlation is not present in the radio quiet sample.}

\item{Of the GRBs we've examined that have very high energy gamma-ray emission (detected by Fermi LAT), eight are from the radio loud sample and only one is from the radio quiet sample.  The VHE emission is extended (lasting a minimum of three times the prompt gamma-ray duration) in all of the radio loud bursts but is {\em not} in the one radio quiet burst with LAT-detected emission.}

\item{There appears to be no difference between jet opening angle and/or the presence (or absence) of a plateau in the afterglow light curves of the radio loud and quiet samples.}
\end{itemize}

  There are many potential progenitor scenarios to explain our results. One promising possibility is that Helium merger systems produce radio loud GRBs and collapsars produce radio quiet GRBs. The dichotomy may also reflect a neutron star vs. black hole central core. Nonetheless, it is clear there is a connection between the inner engine properties (which determine the prompt gamma-ray burst duration) and the environment (which plays a large role in the radio afterglow emission).  
As in the case of AGN, in which after years of debate it became clear there are two distinct populations based on the presence or absence of radio emission, exploring this connection in light of our results may help us better understand the nature of GRB progenitors.

\section*{Acknowledgements}

 We thank the referee for helpful comments and suggestions that led to an improvement of this manuscript.  We also thank Chris Fryer for interesting discussions on progenitor systems.
 M.D. acknowledges the support of the AAS Chretienne International Fellowship.
 BG has received funding from the European Research Council (ERC) under the European Union's Horizon 2020 research and innovation programme (grant agreement no 725246, TEDE, PI Levan).  Work at LANL was done under the auspices of the National Nuclear Security Administration of the US Department of Energy at Los Alamos National Laboratory LA-UR-18-28535.


   





\bibliographystyle{mnras}
\bibliography{refs} 

\begin{thebibliography}{}
\makeatletter
\relax
\def\mn@urlcharsother{\let\do\@makeother \do\$\do\&\do\#\do\^\do\_\do\%\do\~}
\def\mn@doi{\begingroup\mn@urlcharsother \@ifnextchar [ {\mn@doi@}
  {\mn@doi@[]}}
\def\mn@doi@[#1]#2{\def\@tempa{#1}\ifx\@tempa\@empty \href
  {http://dx.doi.org/#2} {doi:#2}\else \href {http://dx.doi.org/#2} {#1}\fi
  \endgroup}
\def\mn@eprint#1#2{\mn@eprint@#1:#2::\@nil}
\def\mn@eprint@arXiv#1{\href {http://arxiv.org/abs/#1} {{\tt arXiv:#1}}}
\def\mn@eprint@dblp#1{\href {http://dblp.uni-trier.de/rec/bibtex/#1.xml}
  {dblp:#1}}
\def\mn@eprint@#1:#2:#3:#4\@nil{\def\@tempa {#1}\def\@tempb {#2}\def\@tempc
  {#3}\ifx \@tempc \@empty \let \@tempc \@tempb \let \@tempb \@tempa \fi \ifx
  \@tempb \@empty \def\@tempb {arXiv}\fi \@ifundefined
  {mn@eprint@\@tempb}{\@tempb:\@tempc}{\expandafter \expandafter \csname
  mn@eprint@\@tempb\endcsname \expandafter{\@tempc}}}

\bibitem[\protect\citeauthoryear{{Abbott} et~al.,}{{Abbott}
  et~al.}{2017}]{Ab17}
{Abbott} B.~P.,  et~al., 2017, \mn@doi [Physical Review Letters]
  {10.1103/PhysRevLett.119.161101}, \href
  {http://adsabs.harvard.edu/abs/2017PhRvL.119p1101A} {119, 161101}

\bibitem[\protect\citeauthoryear{{Aguilera-Dena}, {Langer}, {Moriya}  \&
  {Schootemeijer}}{{Aguilera-Dena} et~al.}{2018}]{Ag18}
{Aguilera-Dena} D.~R.,  {Langer} N.,  {Moriya} T.~J.,   {Schootemeijer} A.,
  2018, \mn@doi [\apj] {10.3847/1538-4357/aabfc1}, \href
  {http://adsabs.harvard.edu/abs/2018ApJ...858..115A} {858, 115}

\bibitem[\protect\citeauthoryear{{Amati} et~al.,}{{Amati} et~al.}{2002}]{Am02}
{Amati} L.,  et~al., 2002, \mn@doi [\aap] {10.1051/0004-6361:20020722}, \href
  {http://adsabs.harvard.edu/abs/2002A%26A...390...81A} {390, 81}

\bibitem[\protect\citeauthoryear{{Berger}}{{Berger}}{2014}]{Berg14}
{Berger} E.,  2014, \mn@doi [\araa] {10.1146/annurev-astro-081913-035926},
  \href {http://adsabs.harvard.edu/abs/2014ARA%26A..52...43B} {52, 43}

\bibitem[\protect\citeauthoryear{{Blanchard}, {Berger}  \& {Fong}}{{Blanchard}
  et~al.}{2016}]{BBF16}
{Blanchard} P.~K.,  {Berger} E.,   {Fong} W.-f.,  2016, \mn@doi [\apj]
  {10.3847/0004-637X/817/2/144}, \href
  {http://adsabs.harvard.edu/abs/2016ApJ...817..144B} {817, 144}

\bibitem[\protect\citeauthoryear{{Blandford} \& {Znajek}}{{Blandford} \&
  {Znajek}}{1977}]{bz77}
{Blandford} R.~D.,  {Znajek} R.~L.,  1977, \mn@doi [\mnras]
  {10.1093/mnras/179.3.433}, \href
  {http://adsabs.harvard.edu/abs/1977MNRAS.179..433B} {179, 433}

\bibitem[\protect\citeauthoryear{{Chandra} \& {Frail}}{{Chandra} \&
  {Frail}}{2012}]{CF12}
{Chandra} P.,  {Frail} D.~A.,  2012, \mn@doi [\apj]
  {10.1088/0004-637X/746/2/156}, \href
  {http://adsabs.harvard.edu/abs/2012ApJ...746..156C} {746, 156}

\bibitem[\protect\citeauthoryear{{Chevalier} \& {Li}}{{Chevalier} \&
  {Li}}{2000}]{CL00}
{Chevalier} R.~A.,  {Li} Z.-Y.,  2000, \mn@doi [\apj] {10.1086/308914}, \href
  {http://adsabs.harvard.edu/abs/2000ApJ...536..195C} {536, 195}

\bibitem[\protect\citeauthoryear{{Chiaberge}, {Gilli}, {Lotz}  \&
  {Norman}}{{Chiaberge} et~al.}{2015}]{Chia15}
{Chiaberge} M.,  {Gilli} R.,  {Lotz} J.~M.,   {Norman} C.,  2015, \mn@doi
  [\apj] {10.1088/0004-637X/806/2/147}, \href
  {http://adsabs.harvard.edu/abs/2015ApJ...806..147C} {806, 147}

\bibitem[\protect\citeauthoryear{{D'Avanzo}}{{D'Avanzo}}{2015}]{DAvanz15}
{D'Avanzo} P.,  2015, \mn@doi [Journal of High Energy Astrophysics]
  {10.1016/j.jheap.2015.07.002}, \href
  {http://adsabs.harvard.edu/abs/2015JHEAp...7...73D} {7, 73}

\bibitem[\protect\citeauthoryear{{Dainotti}, {Cardone}  \&
  {Capozziello}}{{Dainotti} et~al.}{2008}]{Dain08}
{Dainotti} M.~G.,  {Cardone} V.~F.,   {Capozziello} S.,  2008, \mn@doi [\mnras]
  {10.1111/j.1745-3933.2008.00560.x}, \href
  {http://adsabs.harvard.edu/abs/2008MNRAS.391L..79D} {391, L79}

\bibitem[\protect\citeauthoryear{{Dainotti}, {Willingale}, {Capozziello},
  {Fabrizio Cardone}  \& {Ostrowski}}{{Dainotti} et~al.}{2010}]{Dain10}
{Dainotti} M.~G.,  {Willingale} R.,  {Capozziello} S.,  {Fabrizio Cardone} V.,
   {Ostrowski} M.,  2010, \mn@doi [\apjl] {10.1088/2041-8205/722/2/L215}, \href
  {http://adsabs.harvard.edu/abs/2010ApJ...722L.215D} {722, L215}

\bibitem[\protect\citeauthoryear{{Dainotti}, {Petrosian}, {Singal}  \&
  {Ostrowski}}{{Dainotti} et~al.}{2013}]{Dain13}
{Dainotti} M.~G.,  {Petrosian} V.,  {Singal} J.,   {Ostrowski} M.,  2013,
  \mn@doi [\apj] {10.1088/0004-637X/774/2/157}, \href
  {http://adsabs.harvard.edu/abs/2013ApJ...774..157D} {774, 157}

\bibitem[\protect\citeauthoryear{{Dainotti}, {Petrosian}, {Willingale},
  {O'Brien}, {Ostrowski}  \& {Nagataki}}{{Dainotti} et~al.}{2015}]{Dain15}
{Dainotti} M.,  {Petrosian} V.,  {Willingale} R.,  {O'Brien} P.,  {Ostrowski}
  M.,   {Nagataki} S.,  2015, \mn@doi [\mnras] {10.1093/mnras/stv1229}, \href
  {http://adsabs.harvard.edu/abs/2015MNRAS.451.3898D} {451, 3898}

\bibitem[\protect\citeauthoryear{{Dainotti}, {Postnikov}, {Hernandez}  \&
  {Ostrowski}}{{Dainotti} et~al.}{2016}]{Dain16}
{Dainotti} M.~G.,  {Postnikov} S.,  {Hernandez} X.,   {Ostrowski} M.,  2016,
  \mn@doi [\apjl] {10.3847/2041-8205/825/2/L20}, \href
  {http://adsabs.harvard.edu/abs/2016ApJ...825L..20D} {825, L20}

\bibitem[\protect\citeauthoryear{{Dainotti}, {Nagataki}, {Maeda}, {Postnikov}
  \& {Pian}}{{Dainotti} et~al.}{2017a}]{Dain17}
{Dainotti} M.~G.,  {Nagataki} S.,  {Maeda} K.,  {Postnikov} S.,   {Pian} E.,
  2017a, \mn@doi [\aap] {10.1051/0004-6361/201628384}, \href
  {http://adsabs.harvard.edu/abs/2017A%26A...600A..98D} {600, A98}

\bibitem[\protect\citeauthoryear{{Dainotti}, {Hernandez}, {Postnikov},
  {Nagataki}, {O'brien}, {Willingale}  \& {Striegel}}{{Dainotti}
  et~al.}{2017b}]{Dain17c}
{Dainotti} M.~G.,  {Hernandez} X.,  {Postnikov} S.,  {Nagataki} S.,  {O'brien}
  P.,  {Willingale} R.,   {Striegel} S.,  2017b, \mn@doi [\apj]
  {10.3847/1538-4357/aa8a6b}, \href
  {http://adsabs.harvard.edu/abs/2017ApJ...848...88D} {848, 88}

\bibitem[\protect\citeauthoryear{{Dainotti}, {Hernandez}, {Postnikov},
  {Nagataki}, {O'brien}, {Willingale}  \& {Striegel}}{{Dainotti}
  et~al.}{2017c}]{Dain17b}
{Dainotti} M.~G.,  {Hernandez} X.,  {Postnikov} S.,  {Nagataki} S.,  {O'brien}
  P.,  {Willingale} R.,   {Striegel} S.,  2017c, \mn@doi [\apj]
  {10.3847/1538-4357/aa8a6b}, \href
  {http://adsabs.harvard.edu/abs/2017ApJ...848...88D} {848, 88}

\bibitem[\protect\citeauthoryear{{Eichler}, {Livio}, {Piran}  \&
  {Schramm}}{{Eichler} et~al.}{1989}]{Eich89}
{Eichler} D.,  {Livio} M.,  {Piran} T.,   {Schramm} D.~N.,  1989, \mn@doi
  [\nat] {10.1038/340126a0}, \href
  {http://adsabs.harvard.edu/abs/1989Natur.340..126E} {340, 126}

\bibitem[\protect\citeauthoryear{{Fryer} \& {Woosley}}{{Fryer} \&
  {Woosley}}{1998}]{FW98}
{Fryer} C.~L.,  {Woosley} S.~E.,  1998, \mn@doi [\apjl] {10.1086/311493}, \href
  {http://adsabs.harvard.edu/abs/1998ApJ...502L...9F} {502, L9}

\bibitem[\protect\citeauthoryear{{Fryer}, {Belczynski}, {Berger}, {Th{\"o}ne},
  {Ellinger}  \& {Bulik}}{{Fryer} et~al.}{2013}]{Fry13}
{Fryer} C.~L.,  {Belczynski} K.,  {Berger} E.,  {Th{\"o}ne} C.,  {Ellinger} C.,
    {Bulik} T.,  2013, \mn@doi [\apj] {10.1088/0004-637X/764/2/181}, \href
  {http://adsabs.harvard.edu/abs/2013ApJ...764..181F} {764, 181}

\bibitem[\protect\citeauthoryear{{Gao} \& {M{\'e}sz{\'a}ros}}{{Gao} \&
  {M{\'e}sz{\'a}ros}}{2015}]{GM15}
{Gao} H.,  {M{\'e}sz{\'a}ros} P.,  2015, \mn@doi [\apj]
  {10.1088/0004-637X/802/2/90}, \href
  {http://adsabs.harvard.edu/abs/2015ApJ...802...90G} {802, 90}

\bibitem[\protect\citeauthoryear{{Gehrels}, {Ramirez-Ruiz}  \& {Fox}}{{Gehrels}
  et~al.}{2009}]{GRRF09}
{Gehrels} N.,  {Ramirez-Ruiz} E.,   {Fox} D.~B.,  2009, \mn@doi [\araa]
  {10.1146/annurev.astro.46.060407.145147}, \href
  {http://adsabs.harvard.edu/abs/2009ARA%26A..47..567G} {47, 567}

\bibitem[\protect\citeauthoryear{{Ghirlanda}, {Ghisellini}  \&
  {Lazzati}}{{Ghirlanda} et~al.}{2004}]{Ghir04}
{Ghirlanda} G.,  {Ghisellini} G.,   {Lazzati} D.,  2004, \mn@doi [\apj]
  {10.1086/424913}, \href {http://adsabs.harvard.edu/abs/2004ApJ...616..331G}
  {616, 331}

\bibitem[\protect\citeauthoryear{{Gompertz}, {Fruchter}  \& {Pe'er}}{{Gompertz}
  et~al.}{2018}]{GFP18}
{Gompertz} B.~P.,  {Fruchter} A.~S.,   {Pe'er} A.,  2018, preprint, \href
  {http://adsabs.harvard.edu/abs/2018arXiv180207730G} {} (\mn@eprint {arXiv}
  {1802.07730})

\bibitem[\protect\citeauthoryear{{Granot} \& {Sari}}{{Granot} \&
  {Sari}}{2002}]{GS02}
{Granot} J.,  {Sari} R.,  2002, \mn@doi [\apj] {10.1086/338966}, \href
  {http://adsabs.harvard.edu/abs/2002ApJ...568..820G} {568, 820}

\bibitem[\protect\citeauthoryear{{Granot}, {Panaitescu}, {Kumar}  \&
  {Woosley}}{{Granot} et~al.}{2002}]{Gran02}
{Granot} J.,  {Panaitescu} A.,  {Kumar} P.,   {Woosley} S.~E.,  2002, \mn@doi
  [\apjl] {10.1086/340991}, \href
  {http://adsabs.harvard.edu/abs/2002ApJ...570L..61G} {570, L61}

\bibitem[\protect\citeauthoryear{{Hancock}, {Gaensler}  \& {Murphy}}{{Hancock}
  et~al.}{2013}]{HGM13}
{Hancock} P.~J.,  {Gaensler} B.~M.,   {Murphy} T.,  2013, \mn@doi [\apj]
  {10.1088/0004-637X/776/2/106}, \href
  {http://adsabs.harvard.edu/abs/2013ApJ...776..106H} {776, 106}

\bibitem[\protect\citeauthoryear{{Janiuk} \& {Proga}}{{Janiuk} \&
  {Proga}}{2008}]{JP08}
{Janiuk} A.,  {Proga} D.,  2008, \mn@doi [\apj] {10.1086/526511}, \href
  {http://adsabs.harvard.edu/abs/2008ApJ...675..519J} {675, 519}

\bibitem[\protect\citeauthoryear{{Kellermann}, {Condon}, {Kimball}, {Perley}
  \& {Ivezi{\'c}}}{{Kellermann} et~al.}{2016}]{Kell16}
{Kellermann} K.~I.,  {Condon} J.~J.,  {Kimball} A.~E.,  {Perley} R.~A.,
  {Ivezi{\'c}} {\v Z}.,  2016, \mn@doi [\apj] {10.3847/0004-637X/831/2/168},
  \href {http://adsabs.harvard.edu/abs/2016ApJ...831..168K} {831, 168}

\bibitem[\protect\citeauthoryear{{Kocevski} \& {Petrosian}}{{Kocevski} \&
  {Petrosian}}{2013}]{KP13}
{Kocevski} D.,  {Petrosian} V.,  2013, \mn@doi [\apj]
  {10.1088/0004-637X/765/2/116}, \href
  {http://adsabs.harvard.edu/abs/2013ApJ...765..116K} {765, 116}

\bibitem[\protect\citeauthoryear{{Kumar} \& {Zhang}}{{Kumar} \&
  {Zhang}}{2015}]{KZ15}
{Kumar} P.,  {Zhang} B.,  2015, \mn@doi [\physrep]
  {10.1016/j.physrep.2014.09.008}, \href
  {http://adsabs.harvard.edu/abs/2015PhR...561....1K} {561, 1}

\bibitem[\protect\citeauthoryear{{Lan} et~al.,}{{Lan} et~al.}{2018}]{Lan18}
{Lan} L.,  et~al., 2018, preprint, \href
  {http://adsabs.harvard.edu/abs/2018arXiv180606690L} {} (\mn@eprint {arXiv}
  {1806.06690})

\bibitem[\protect\citeauthoryear{{Laskar} et~al.,}{{Laskar}
  et~al.}{2013}]{Las13}
{Laskar} T.,  et~al., 2013, \mn@doi [\apj] {10.1088/0004-637X/776/2/119}, \href
  {http://adsabs.harvard.edu/abs/2013ApJ...776..119L} {776, 119}

\bibitem[\protect\citeauthoryear{{Laskar} et~al.,}{{Laskar}
  et~al.}{2016}]{Las16}
{Laskar} T.,  et~al., 2016, \mn@doi [\apj] {10.3847/1538-4357/833/1/88}, \href
  {http://adsabs.harvard.edu/abs/2016ApJ...833...88L} {833, 88}

\bibitem[\protect\citeauthoryear{{Laskar} et~al.,}{{Laskar}
  et~al.}{2018}]{Las18}
{Laskar} T.,  et~al., 2018, \mn@doi [\apj] {10.3847/1538-4357/aacbcc}, \href
  {http://adsabs.harvard.edu/abs/2018ApJ...862...94L} {862, 94}

\bibitem[\protect\citeauthoryear{{Levan}, {Crowther}, {de Grijs}, {Langer},
  {Xu}  \& {Yoon}}{{Levan} et~al.}{2016}]{Lev16}
{Levan} A.,  {Crowther} P.,  {de Grijs} R.,  {Langer} N.,  {Xu} D.,   {Yoon}
  S.-C.,  2016, \mn@doi [\ssr] {10.1007/s11214-016-0312-x}, \href
  {http://adsabs.harvard.edu/abs/2016SSRv..202...33L} {202, 33}

\bibitem[\protect\citeauthoryear{{Littlejohns}, {Tanvir}, {Willingale},
  {Evans}, {O'Brien}  \& {Levan}}{{Littlejohns} et~al.}{2013}]{LJ13}
{Littlejohns} O.~M.,  {Tanvir} N.~R.,  {Willingale} R.,  {Evans} P.~A.,
  {O'Brien} P.~T.,   {Levan} A.~J.,  2013, \mn@doi [\mnras]
  {10.1093/mnras/stt1841}, \href
  {http://adsabs.harvard.edu/abs/2013MNRAS.436.3640L} {436, 3640}

\bibitem[\protect\citeauthoryear{{Lloyd-Ronning} \& {Fryer}}{{Lloyd-Ronning} \&
  {Fryer}}{2017}]{LRF17}
{Lloyd-Ronning} N.~M.,  {Fryer} C.~L.,  2017, \mn@doi [\mnras]
  {10.1093/mnras/stx313}, \href
  {http://adsabs.harvard.edu/abs/2017MNRAS.467.3413L} {467, 3413}

\bibitem[\protect\citeauthoryear{{Lloyd-Ronning} \&
  {Petrosian}}{{Lloyd-Ronning} \& {Petrosian}}{2002}]{LP02}
{Lloyd-Ronning} N.~M.,  {Petrosian} V.,  2002, \mn@doi [\apj] {10.1086/324484},
  \href {http://adsabs.harvard.edu/abs/2002ApJ...565..182L} {565, 182}

\bibitem[\protect\citeauthoryear{{Lloyd-Ronning}, {Lei}  \&
  {Xie}}{{Lloyd-Ronning} et~al.}{2018}]{LR18}
{Lloyd-Ronning} N.,  {Lei} W.-h.,   {Xie} W.,  2018, \mn@doi [\mnras]
  {10.1093/mnras/sty1030}, \href
  {http://adsabs.harvard.edu/abs/2018MNRAS.478.3525L} {478, 3525}

\bibitem[\protect\citeauthoryear{{Lloyd}, {Petrosian}  \& {Mallozzi}}{{Lloyd}
  et~al.}{2000}]{LPM00}
{Lloyd} N.~M.,  {Petrosian} V.,   {Mallozzi} R.~S.,  2000, \mn@doi [\apj]
  {10.1086/308742}, \href {http://adsabs.harvard.edu/abs/2000ApJ...534..227L}
  {534, 227}

\bibitem[\protect\citeauthoryear{{MacFadyen} \& {Woosley}}{{MacFadyen} \&
  {Woosley}}{1999}]{MW99}
{MacFadyen} A.~I.,  {Woosley} S.~E.,  1999, \mn@doi [\apj] {10.1086/307790},
  \href {http://adsabs.harvard.edu/abs/1999ApJ...524..262M} {524, 262}

\bibitem[\protect\citeauthoryear{{Matsumoto}, {Nakar}  \& {Piran}}{{Matsumoto}
  et~al.}{2018}]{Mat18}
{Matsumoto} T.,  {Nakar} E.,   {Piran} T.,  2018, preprint, \href
  {http://adsabs.harvard.edu/abs/2018arXiv180704756M} {} (\mn@eprint {arXiv}
  {1807.04756})

\bibitem[\protect\citeauthoryear{{Mazzali}, {McFadyen}, {Woosley}, {Pian}  \&
  {Tanaka}}{{Mazzali} et~al.}{2014}]{Mazz14}
{Mazzali} P.~A.,  {McFadyen} A.~I.,  {Woosley} S.~E.,  {Pian} E.,   {Tanaka}
  M.,  2014, \mn@doi [\mnras] {10.1093/mnras/stu1124}, \href
  {http://adsabs.harvard.edu/abs/2014MNRAS.443...67M} {443, 67}

\bibitem[\protect\citeauthoryear{{M{\'e}sz{\'a}ros}}{{M{\'e}sz{\'a}ros}}{2006}]{Mesz06}
{M{\'e}sz{\'a}ros} P.,  2006, \mn@doi [Reports on Progress in Physics]
  {10.1088/0034-4885/69/8/R01}, \href
  {http://adsabs.harvard.edu/abs/2006RPPh...69.2259M} {69, 2259}

\bibitem[\protect\citeauthoryear{{M{\'e}sz{\'a}ros}, {Rees}  \&
  {Wijers}}{{M{\'e}sz{\'a}ros} et~al.}{1998}]{MRW98}
{M{\'e}sz{\'a}ros} P.,  {Rees} M.~J.,   {Wijers} R.~A.~M.~J.,  1998, \mn@doi
  [\apj] {10.1086/305635}, \href
  {http://adsabs.harvard.edu/abs/1998ApJ...499..301M} {499, 301}

\bibitem[\protect\citeauthoryear{{Metzger}, {Margalit}, {Kasen}  \&
  {Quataert}}{{Metzger} et~al.}{2015}]{Metz15}
{Metzger} B.~D.,  {Margalit} B.,  {Kasen} D.,   {Quataert} E.,  2015, \mn@doi
  [\mnras] {10.1093/mnras/stv2224}, \href
  {http://adsabs.harvard.edu/abs/2015MNRAS.454.3311M} {454, 3311}

\bibitem[\protect\citeauthoryear{{Nava}}{{Nava}}{2018}]{Nava18}
{Nava} L.,  2018, preprint, \href
  {http://adsabs.harvard.edu/abs/2018arXiv180401524N} {} (\mn@eprint {arXiv}
  {1804.01524})

\bibitem[\protect\citeauthoryear{{Nemmen}, {Georganopoulos}, {Guiriec},
  {Meyer}, {Gehrels}  \& {Sambruna}}{{Nemmen} et~al.}{2012}]{Nem12}
{Nemmen} R.~S.,  {Georganopoulos} M.,  {Guiriec} S.,  {Meyer} E.~T.,  {Gehrels}
  N.,   {Sambruna} R.~M.,  2012, \mn@doi [Science] {10.1126/science.1227416},
  \href {http://adsabs.harvard.edu/abs/2012Sci...338.1445N} {338, 1445}

\bibitem[\protect\citeauthoryear{{O'Brien} et~al.,}{{O'Brien}
  et~al.}{2006}]{Obri06}
{O'Brien} P.~T.,  et~al., 2006, \mn@doi [\apj] {10.1086/505457}, \href
  {http://adsabs.harvard.edu/abs/2006ApJ...647.1213O} {647, 1213}

\bibitem[\protect\citeauthoryear{{Panaitescu} \& {Vestrand}}{{Panaitescu} \&
  {Vestrand}}{2008}]{PV08}
{Panaitescu} A.,  {Vestrand} W.~T.,  2008, \mn@doi [\mnras]
  {10.1111/j.1365-2966.2008.13231.x}, \href
  {http://adsabs.harvard.edu/abs/2008MNRAS.387..497P} {387, 497}

\bibitem[\protect\citeauthoryear{{Pescalli} et~al.,}{{Pescalli}
  et~al.}{2016}]{Pesc16}
{Pescalli} A.,  et~al., 2016, \mn@doi [\aap] {10.1051/0004-6361/201526760},
  \href {http://adsabs.harvard.edu/abs/2016A%26A...587A..40P} {587, A40}

\bibitem[\protect\citeauthoryear{{Piran}}{{Piran}}{2004}]{Pir04}
{Piran} T.,  2004, \mn@doi [Reviews of Modern Physics]
  {10.1103/RevModPhys.76.1143}, \href
  {http://adsabs.harvard.edu/abs/2004RvMP...76.1143P} {76, 1143}

\bibitem[\protect\citeauthoryear{{Qin} et~al.,}{{Qin} et~al.}{2013}]{Qin13}
{Qin} Y.,  et~al., 2013, \mn@doi [\apj] {10.1088/0004-637X/763/1/15}, \href
  {http://adsabs.harvard.edu/abs/2013ApJ...763...15Q} {763, 15}

\bibitem[\protect\citeauthoryear{{Rea}, {Gull{\'o}n}, {Pons}, {Perna},
  {Dainotti}, {Miralles}  \& {Torres}}{{Rea} et~al.}{2015}]{Rea15}
{Rea} N.,  {Gull{\'o}n} M.,  {Pons} J.~A.,  {Perna} R.,  {Dainotti} M.~G.,
  {Miralles} J.~A.,   {Torres} D.~F.,  2015, \mn@doi [\apj]
  {10.1088/0004-637X/813/2/92}, \href
  {http://adsabs.harvard.edu/abs/2015ApJ...813...92R} {813, 92}

\bibitem[\protect\citeauthoryear{{Rossi}, {Lazzati}  \& {Rees}}{{Rossi}
  et~al.}{2002}]{RLR02}
{Rossi} E.,  {Lazzati} D.,   {Rees} M.~J.,  2002, \mn@doi [\mnras]
  {10.1046/j.1365-8711.2002.05363.x}, \href
  {http://adsabs.harvard.edu/abs/2002MNRAS.332..945R} {332, 945}

\bibitem[\protect\citeauthoryear{{Rowlinson}, {Gompertz}, {Dainotti},
  {O'Brien}, {Wijers}  \& {van der Horst}}{{Rowlinson} et~al.}{2014}]{Row14}
{Rowlinson} A.,  {Gompertz} B.~P.,  {Dainotti} M.,  {O'Brien} P.~T.,  {Wijers}
  R.~A.~M.~J.,   {van der Horst} A.~J.,  2014, \mn@doi [\mnras]
  {10.1093/mnras/stu1277}, \href
  {http://adsabs.harvard.edu/abs/2014MNRAS.443.1779R} {443, 1779}

\bibitem[\protect\citeauthoryear{{Sari}, {Piran}  \& {Narayan}}{{Sari}
  et~al.}{1998}]{SPN98}
{Sari} R.,  {Piran} T.,   {Narayan} R.,  1998, \mn@doi [\apjl]
  {10.1086/311269}, \href {http://adsabs.harvard.edu/abs/1998ApJ...497L..17S}
  {497, L17}

\bibitem[\protect\citeauthoryear{{Si} et~al.,}{{Si} et~al.}{2018}]{Si18}
{Si} S.-K.,  et~al., 2018, preprint, \href
  {http://adsabs.harvard.edu/abs/2018arXiv180700241S} {} (\mn@eprint {arXiv}
  {1807.00241})

\bibitem[\protect\citeauthoryear{{Stratta}, {Dainotti}, {Dall'Osso},
  {Hernandez}  \& {De Cesare}}{{Stratta} et~al.}{2018}]{SD18}
{Stratta} G.,  {Dainotti} M.~G.,  {Dall'Osso} S.,  {Hernandez} X.,   {De
  Cesare} G.,  2018, preprint, \href
  {http://adsabs.harvard.edu/abs/2018arXiv180408652S} {} (\mn@eprint {arXiv}
  {1804.08652})

\bibitem[\protect\citeauthoryear{{The Fermi LAT Collaboration}}{{The Fermi LAT
  Collaboration}}{2018}]{Ferm18}
{The Fermi LAT Collaboration} 2018, preprint, \href
  {http://adsabs.harvard.edu/abs/2018arXiv180801683T} {} (\mn@eprint {arXiv}
  {1808.01683})

\bibitem[\protect\citeauthoryear{{Thompson}}{{Thompson}}{1994}]{Thomp94}
{Thompson} C.,  1994, \mn@doi [\mnras] {10.1093/mnras/270.3.480}, \href
  {http://adsabs.harvard.edu/abs/1994MNRAS.270..480T} {270, 480}

\bibitem[\protect\citeauthoryear{{Toma}, {Yoon}  \& {Bromm}}{{Toma}
  et~al.}{2016}]{TYB16}
{Toma} K.,  {Yoon} S.-C.,   {Bromm} V.,  2016, \mn@doi [\ssr]
  {10.1007/s11214-016-0250-7}, \href
  {http://adsabs.harvard.edu/abs/2016SSRv..202..159T} {202, 159}

\bibitem[\protect\citeauthoryear{{Usov}}{{Usov}}{1992}]{Usov92}
{Usov} V.~V.,  1992, \mn@doi [\nat] {10.1038/357472a0}, \href
  {http://adsabs.harvard.edu/abs/1992Natur.357..472U} {357, 472}

\bibitem[\protect\citeauthoryear{{Wang}, {Zhang}, {Liang}, {Lu}, {Lin}, {Li}
  \& {Li}}{{Wang} et~al.}{2018}]{Wang18}
{Wang} X.-G.,  {Zhang} B.,  {Liang} E.-W.,  {Lu} R.-J.,  {Lin} D.-B.,  {Li} J.,
    {Li} L.,  2018, \mn@doi [\apj] {10.3847/1538-4357/aabc13}, \href
  {http://adsabs.harvard.edu/abs/2018ApJ...859..160W} {859, 160}

\bibitem[\protect\citeauthoryear{{Willingale} et~al.,}{{Willingale}
  et~al.}{2007}]{Will07}
{Willingale} R.,  et~al., 2007, \mn@doi [\apj] {10.1086/517989}, \href
  {http://adsabs.harvard.edu/abs/2007ApJ...662.1093W} {662, 1093}

\bibitem[\protect\citeauthoryear{{Wilson} \& {Colbert}}{{Wilson} \&
  {Colbert}}{1995}]{WC95}
{Wilson} A.~S.,  {Colbert} E.~J.~M.,  1995, \mn@doi [\apj] {10.1086/175054},
  \href {http://adsabs.harvard.edu/abs/1995ApJ...438...62W} {438, 62}

\bibitem[\protect\citeauthoryear{{Woosley}}{{Woosley}}{1993}]{Woos93}
{Woosley} S.~E.,  1993, \mn@doi [\apj] {10.1086/172359}, \href
  {http://adsabs.harvard.edu/abs/1993ApJ...405..273W} {405, 273}

\bibitem[\protect\citeauthoryear{{Wu}, {Zhang}, {Lei}, {Zou}, {Liang}  \&
  {Cao}}{{Wu} et~al.}{2016}]{wu16}
{Wu} Q.,  {Zhang} B.,  {Lei} W.-H.,  {Zou} Y.-C.,  {Liang} E.-W.,   {Cao} X.,
  2016, \mn@doi [\mnras] {10.1093/mnrasl/slv136}, \href
  {http://adsabs.harvard.edu/abs/2016MNRAS.455L...1W} {455, L1}

\bibitem[\protect\citeauthoryear{{Zaninoni}, {Bernardini}, {Margutti}, {Oates}
  \& {Chincarini}}{{Zaninoni} et~al.}{2013}]{Zan13}
{Zaninoni} E.,  {Bernardini} M.~G.,  {Margutti} R.,  {Oates} S.,   {Chincarini}
  G.,  2013, \mn@doi [\aap] {10.1051/0004-6361/201321221}, \href
  {http://adsabs.harvard.edu/abs/2013A%26A...557A..12Z} {557, A12}

\bibitem[\protect\citeauthoryear{{Zhang} \& {Fryer}}{{Zhang} \&
  {Fryer}}{2001}]{ZF00}
{Zhang} W.,  {Fryer} C.~L.,  2001, \mn@doi [\apj] {10.1086/319734}, \href
  {http://adsabs.harvard.edu/abs/2001ApJ...550..357Z} {550, 357}

\bibitem[\protect\citeauthoryear{{Zhang} \& {M{\'e}sz{\'a}ros}}{{Zhang} \&
  {M{\'e}sz{\'a}ros}}{2001}]{Zhang01}
{Zhang} B.,  {M{\'e}sz{\'a}ros} P.,  2001, \mn@doi [\apjl] {10.1086/320255},
  \href {http://adsabs.harvard.edu/abs/2001ApJ...552L..35Z} {552, L35}

\bibitem[\protect\citeauthoryear{{Zhang} \& {M{\'e}sz{\'a}ros}}{{Zhang} \&
  {M{\'e}sz{\'a}ros}}{2004}]{ZM04}
{Zhang} B.,  {M{\'e}sz{\'a}ros} P.,  2004, \mn@doi [International Journal of
  Modern Physics A] {10.1142/S0217751X0401746X}, \href
  {http://adsabs.harvard.edu/abs/2004IJMPA..19.2385Z} {19, 2385}

\bibitem[\protect\citeauthoryear{{Zhang}, {Liang}, {Sun}, {Zhang}, {Lu}  \&
  {Zhang}}{{Zhang} et~al.}{2013}]{Zhang13}
{Zhang} J.,  {Liang} E.-W.,  {Sun} X.-N.,  {Zhang} B.,  {Lu} Y.,   {Zhang}
  S.-N.,  2013, \mn@doi [\apjl] {10.1088/2041-8205/774/1/L5}, \href
  {http://adsabs.harvard.edu/abs/2013ApJ...774L...5Z} {774, L5}

\makeatother
\end{thebibliography}

\end{document}